\documentclass[aps,prb,superscriptaddress,twocolumn,longbibliography]{revtex4-2}
\usepackage{bbm}
\usepackage{graphicx}
\usepackage{dcolumn}
\usepackage{bm}
\usepackage{subfigure}
\usepackage{amsmath}
\usepackage{feynmf}
\usepackage[colorlinks=true,linkcolor=blue,anchorcolor=blue, citecolor=cyan,urlcolor=cyan]{hyperref}
\usepackage[mathlines]{lineno}
\usepackage{ulem}
\begin{document}
	\title{Direction-selective triplet pairing and spin-edge locking in altermagnetic metals}
	\author{Lie Yuan}
	\author{Junkang Huang}

	\affiliation{Guangdong Basic Research Center of Excellence for Structure and Fundamental Interactions of Matter, Guangdong Provincial Key Laboratory of Quantum Engineering and Quantum Materials, School of Physics, South China Normal University, Guangzhou 510006, China}
		\author{Yu-Xuan Li}
	\email{yxliphy@gmail.com}
	\author{Tao Zhou}
	\email{tzhou@scnu.edu.cn}
	\affiliation{Guangdong Basic Research Center of Excellence for Structure and Fundamental Interactions of Matter, Guangdong Provincial Key Laboratory of Quantum Engineering and Quantum Materials, School of Physics, South China Normal University, Guangzhou 510006, China}
	\affiliation{Guangdong-Hong Kong Joint Laboratory of Quantum Matter, Frontier Research Institute for Physics, South China Normal University, Guangzhou 510006, China}

	\begin{abstract}
We investigate self-consistent unconventional superconductivity in a two-dimensional $d$-wave altermagnetic metal.
We find that momentum-dependent altermagnetic spin splitting suppresses opposite-spin singlet pairing and stabilizes highly anisotropic equal-spin triplet order. 
In the spin-conserving limit, this directional triplet pairing gives rise to nearly dispersionless Majorana boundary states associated with effective one-dimensional topological channels. 
Rashba spin-orbit coupling mixes spin sectors, activates additional pairing components, and drives the system into a mixed-parity superconducting state with dispersive Majorana boundary states. 
The spin-resolved boundary spectra further reveal a characteristic locking between boundary orientation and spin polarization, reflecting the underlying altermagnetic symmetry. 
These results identify altermagnetic spin splitting as an intrinsic mechanism for selecting unconventional pairing and generating spin-resolved Majorana boundary states without external magnetic fields.
	\end{abstract}
	\maketitle
	\section{Introduction}
	
	The emergence of unconventional quantum phases in correlated and topological materials often arises from the interplay of multiple electronic degrees of freedom, including charge, spin, orbital, and superconducting orders~\cite{bergeretOdd2005,bergStriped2009,scalapinoCommon2012,fradkinColloquium2015,scalapinoWave1986}. 
	Among these, the interplay between magnetism and superconductivity has attracted extensive attention~\cite{dikinCoexistence2011,liCoexistence2011}. 
	Magnetic order generally suppresses conventional spin-singlet superconductivity through pair-breaking effects, but it can also reconstruct the spin and momentum structure of quasiparticles and thereby promote unconventional pairing~\cite{bergeretOdd2005}. 
	Understanding how magnetism suppresses, selects, or stabilizes superconducting order is therefore central to the search for novel superconducting and topological phases~\cite{sigristPhenomenological1991,alicea_new_2012,qi_topological_2011,flensberg_engineered_2021}.
	
	Spin splitting plays a central role in realizing unconventional and topological superconductivity, as it lifts spin degeneracy~\cite{alicea_new_2012,qi_topological_2011,sato_topological_2017}. 
	When combined with spin-orbit coupling, spin splitting can generate effective $p$-wave pairing and thereby support Majorana excitations~\cite{alicea_new_2012,beenakker_search_2013}. 
	In conventional platforms, spin splitting is typically produced by external Zeeman fields or ferromagnetic order~\cite{Ranran_SuperSpintronics_2023,flensberg_engineered_2021,Das_Rashba_2026}. 
	However, these routes often introduce orbital and paramagnetic depairing, as well as stray-field effects, which are detrimental to superconductivity~\cite{sigristPhenomenological1991,bergeretOdd2005}. 
	This motivates the search for magnetic systems in which intrinsic spin splitting can coexist with superconductivity.
	
	Altermagnets have recently emerged as a distinct class of collinear compensated magnets beyond the conventional ferromagnet-antiferromagnet dichotomy~\cite{smejkal_emerging_2022,mazin_editorial_2022,xiao_spin_2024,bai_altermagnetism_2024,fender_altermagnetism_2025,liu_altermagnetism_2025,Jungwirth2026,Qihang_Rise_2026,Luo_Symmetry_2026}. 
	Unlike conventional antiferromagnets, altermagnets can host momentum-dependent spin splitting despite vanishing net magnetization and, in many cases, even the absence of relativistic spin-orbit coupling~\cite{Libor_Beyond_2022,smejkal_emerging_2022}. 
	This spin splitting originates from crystalline symmetries, where opposite-spin sublattices are related by time reversal combined with rotation or mirror operations rather than by inversion or translation~\cite{Libor_Beyond_2022}. 
Consequently, altermagnets exhibit anisotropic spin-polarized Fermi surfaces that are qualitatively distinct from those generated by Zeeman fields or ferromagnets. 
	This makes them a natural platform for exploring unconventional superconductivity in the absence of macroscopic magnetization~\cite{liu_altermagnetism_2025}.
	
	Superconductivity in altermagnetic systems has therefore attracted increasing attention~\cite{fukaya_superconducting_2025,liu_altermagnetism_2025,ouassou_dc_2023,Papaj_2023,Sun_Andreev_2023,zhu_topological_2023,ghorashi_altermagnetic_2024,ezawa_topological_2024,parshukov_exotic_2025,Heinsdorf_proxi_2026,Alam_prox_2026,xiao_nodal_2026,Zou_Intertwined_2026,de_carvalho_unconventional_2024,chakraborty_constraints_2025,Beenakker_Andreev_2023,li_majorana_2023,Zhu_Dislocation_2024,zhang_finite-momentum_2024,lu_ensuremathvarphi_2024,cheng_orientation-dependent_2024,li_realizing_2024,gondolf_local_2025,Qiao_MCM_2025,alipourzadeh_andreev_2025,hu_nonlinear_2025,Law_Pseudo_2025,mondal_distinguishing_2025,hadjipaschalisMajoranas2025,Pal_jose_2025,maeda_classification_2025,Chuangli_2026,Dsouza_oddmagne_2026,Sharma_Doublepeak_2026,Sun_Majorana_2026,Franz_Persistent_2026,Wan_Floqu_2026,Hu_FFLO_2026,Mazanik_Probe_2026}. 
	Previous studies have explored both intrinsic pairing instabilities in altermagnetic metals and proximity-induced superconducting phenomena in altermagnet-superconductor hybrid structures. 
	These works have revealed a variety of unconventional responses, including equal-spin triplet pairing, mixed singlet-triplet pairing, finite-momentum superconductivity, anomalous Josephson and Andreev effects, and possible routes to topological superconductivity~\cite{liu_altermagnetism_2025,zhu_topological_2023,rasmussen_inherent_2025,parshukov_exotic_2025,Heinsdorf_proxi_2026,Alam_prox_2026,Zou_Intertwined_2026,ouassou_dc_2023,Sun_Andreev_2023,Papaj_2023,ghorashi_altermagnetic_2024,ezawa_topological_2024,xiao_nodal_2026}. 
	More recent self-consistent calculations have further indicated that $d$-wave altermagnetism, especially when combined with Rashba spin-orbit coupling (RSOC), can support mixed-parity superconducting states~\cite{de_carvalho_unconventional_2024,rasmussen_inherent_2025,chakraborty_constraints_2025,Zou_Intertwined_2026}. 
	
	Despite these advances, the microscopic mechanism by which altermagnetic spin splitting selects superconducting order remains unclear. 
	When singlet and triplet pairing channels compete, a central question is whether the superconducting state is imposed by construction or emerges self-consistently. 
A further key issue is how the anisotropy dictated by altermagnetic symmetry propagates from the normal‑state electronic structure to the pairing order, and ultimately to Majorana boundary states and their spin‑resolved spectral signatures.

    In this work, we develop a minimal self-consistent framework for superconductivity in a two-dimensional $d$-wave altermagnetic metal with RSOC and short-range attractive interactions. 
	The pairing amplitudes are determined dynamically, allowing the opposite-spin singlet and equal-spin triplet channels to compete without imposing any predefined pairing symmetry. 
	This approach reveals that the $d_{x^2-y^2}$ altermagnetic spin splitting acts as a symmetry-selective mechanism: it suppresses singlet pairing, generates a quadratic anisotropy between the $p_x$ and $p_y$ triplet channels, and selects direction-dependent equal-spin triplet components. 
	The preserved $\mathcal{C}_{4z}\mathcal{T}$ symmetry then relates the dominant triplet components in opposite spin sectors. 
	RSOC relaxes this spin-selective structure by mixing spin sectors, thereby promoting a mixed-parity superconducting state. 
	The resulting superconducting state supports Majorana boundary states, whose spin-resolved spectral weights reveal a characteristic spin-edge locking dictated by the altermagnetic symmetry.
	
	The remainder of this paper is organized as follows. 
	In Sec.~\ref{sec:model}, we introduce the minimal model for a two-dimensional $d$-wave altermagnetic metal with RSOC and outline the self-consistent mean-field framework for determining the superconducting order parameters. 
	In Sec.~\ref{sec:results}, we present the pairing phase diagram, analyze the altermagnetism-driven anisotropy in the singlet-triplet competition, and discuss the resulting boundary spectra, including Majorana modes and their spin-resolved structure. 
	Finally, Sec.~\ref{sec:summary} summarizes the main results and discusses their broader implications.

	\section{Model and Method}\label{sec:model}
	We consider an effective two-band model for a two-dimensional $d$-wave altermagnetic metal with RSOC. 
	This model can be viewed as the low-energy projection of a four-band altermagnetic system in the strong spin-sublattice locking regime, where the bands crossing the Fermi level are described by an effective spin degree of freedom. 
	The normal-state Hamiltonian is $	H_{\rm N}
	=
	\sum_{\boldsymbol{k}}
	c_{\boldsymbol{k}}^\dagger
	h_0(\boldsymbol{k})
	c_{\boldsymbol{k}},$
	where  $c_{\boldsymbol{k}}^\dagger=(c_{\boldsymbol{k}\uparrow}^\dagger,c_{\boldsymbol{k}\downarrow}^\dagger)$, and the matrix $h_0(\boldsymbol{k})$ is given by
	\begin{equation}
		\begin{aligned}
			h_0(\boldsymbol{k})
			=&
			\epsilon(\boldsymbol{k})\sigma_0
			+
			2J_0(\cos k_x-\cos k_y)\sigma_z
			\\
			&+
			2\lambda(\sin k_y\sigma_x-\sin k_x\sigma_y).
		\end{aligned}
		\label{eq:normal_hamiltonian}
	\end{equation}
	Here $	\epsilon(\boldsymbol{k})
	=
	-2t_0(\cos k_x+\cos k_y)-\mu ,$ where $t_0$ is the nearest-neighbor hopping amplitude and $\mu$ is the chemical potential. 
	The second term in Eq.~\eqref{eq:normal_hamiltonian} describes the $d_{x^2-y^2}$-type altermagnetic exchange field. Its form factor $\cos k_x-\cos k_y$ produces momentum-dependent spin splitting with opposite signs in different regions of the Brillouin zone. 
	This term breaks $\mathcal{T}$ and $\mathcal{C}_{4z}$ separately, but preserves the combined $\mathcal{C}_{4z}\mathcal{T}$ symmetry, leading to spin-split yet compensated bands. 
	The last term is the RSOC induced by inversion-symmetry breaking, which mixes the two spin sectors.
	
We consider superconductivity arising from magnetic exchange interactions on nearest-neighbor bonds.  The corresponding interaction Hamiltonian can be written as
	\begin{equation}
	H_I =-V\sum_{\langle{ ij}\rangle}\sum_{\sigma,\sigma^\prime}   c_{i\sigma}^{\dagger} c_{{j}\sigma^\prime}^{\dagger}c_{{ i}\sigma^\prime} c_{{ j}\sigma} ,
	\end{equation}
	where $V$ denotes the pairing strength. The cases $\sigma \neq \sigma'$ and $\sigma = \sigma'$ correspond to spin-singlet and equal-spin triplet pairings, respectively. At the mean-field level, we decouple this interaction to obtain the following bond order parameters:
$	\Delta^{s}_{i,j}
		=
		\frac{V}{2}
		\left\langle
		c_{i\uparrow}c_{j\downarrow}
		-
		c_{i\downarrow}c_{j\uparrow}
		\right\rangle $ 
and
$		\Delta^{t,\sigma}_{i,j}
		=
		V
		\left\langle
		c_{i\sigma}c_{j\sigma}
		\right\rangle$.
The symmetry properties under bond reversal are: 
\begin{subequations}
\begin{align}
\Delta_{i,i+\hat{\mu}}^{s} &= +\Delta_{i+\hat{\mu},i}^{s}, \\
\Delta_{i,i+\hat{\mu}}^{t,\sigma} &= -\Delta_{i+\hat{\mu},i}^{t,\sigma},
\end{align}
\end{subequations}
where $\mu = x,y$ denotes the bond direction. The plus sign for singlet pairing reflects its even orbital parity, while the minus sign for triplet pairing reflects its odd parity.

At the mean-field level, the magnitude of the singlet and triplet order parameters along the $\mu$-direction can be expressed as,
\begin{subequations}
\begin{align}
\Delta_{\mu}^{s} &= \frac{V}{2N} \sum_{\boldsymbol{k}} \cos k_{\mu} \, \mathcal{F}_{\uparrow\downarrow}(\boldsymbol{k}), \label{OP:singlet} \\
\Delta_{\mu}^{t,\sigma} &= \frac{V}{2N} \sum_{\boldsymbol{k}} i \sin k_{\mu} \, \mathcal{F}_{\sigma\sigma}(\boldsymbol{k}), \label{OP:triplet}
\end{align}
\end{subequations}
where $N$ is the number of lattice sites and $\mathcal{F}_{\sigma\sigma'}(\boldsymbol{k}) = \langle c_{\boldsymbol{k}\sigma} c_{-\boldsymbol{k}\sigma'}\rangle$ is the anomalous expectation value.

We introduce the Nambu spinor $\Psi_{\boldsymbol{k}} = (c_{\boldsymbol{k}\uparrow}, c_{\boldsymbol{k}\downarrow}, c_{-\boldsymbol{k}\uparrow}^{\dagger}, c_{-\boldsymbol{k}\downarrow}^{\dagger})^{T}$. The mean-field Bogoliubov–de Gennes (BdG) Hamiltonian is $H_{\mathrm{BdG}} = \frac{1}{2}\sum_{\boldsymbol{k}} \Psi_{\boldsymbol{k}}^{\dagger} \mathcal{H}_{\mathrm{BdG}}(\boldsymbol{k}) \Psi_{\boldsymbol{k}}$, with
\begin{equation}
\mathcal{H}_{\mathrm{BdG}}(\boldsymbol{k}) = 
\begin{pmatrix}
h_0(\boldsymbol{k}) & \Delta(\boldsymbol{k}) \\
\Delta^{\dagger}(\boldsymbol{k}) & -h_0^{T}(-\boldsymbol{k})
\end{pmatrix},
\label{eq:BdG}
\end{equation}
where $\Delta(\boldsymbol{k})$ is the pairing matrix:
\begin{equation}
\Delta(\boldsymbol{k}) = 
\begin{pmatrix}
\Delta^{t,\uparrow}(\boldsymbol{k}) & \Delta^{s}(\boldsymbol{k}) \\
-\Delta^{s}(\boldsymbol{k}) & \Delta^{t,\downarrow}(\boldsymbol{k})
\end{pmatrix},
\label{eq:pairing_matrix}
\end{equation}
and the pairing functions in the singlet and triplet channels are expressed as
\begin{subequations}
\begin{align}
\Delta^{s}(\boldsymbol{k}) &= 2\Delta_{x}^{s}\cos k_x + 2\Delta_{y}^{s}\cos k_y, \label{eq:singlet_k} \\
\Delta^{t,\sigma}(\boldsymbol{k}) &= 2i\Bigl( \Delta_{x}^{t,\sigma}\sin k_x + \Delta_{y}^{t,\sigma}\sin k_y \Bigr). \label{eq:triplet_k}
\end{align}
\end{subequations}

Diagonalizing $\mathcal{H}_{\mathrm{BdG}}(\boldsymbol{k})$ gives the quasiparticle eigenvalues $E_{\boldsymbol{k}n}$ and eigenvectors
$\Phi_{\boldsymbol{k}n} = \bigl( u_{1\boldsymbol{k}n},\; u_{2\boldsymbol{k}n},\; u_{3\boldsymbol{k}n},\; u_{4\boldsymbol{k}n} \bigr)^T$,
where the four components correspond to $(c_{\boldsymbol{k}\uparrow}, c_{\boldsymbol{k}\downarrow}, c_{-\boldsymbol{k}\uparrow}^{\dagger}, c_{-\boldsymbol{k}\downarrow}^{\dagger})$. Using Eqs.~\eqref{OP:singlet} and \eqref{OP:triplet}, the order parameter magnitudes along the $\mu$ direction can be expressed as
\begin{equation}
\begin{aligned}
\Delta_{\mu}^{s}
&= \frac{V}{2N} \sum_{\boldsymbol{k},n} u_{1\boldsymbol{k}n} u_{4\boldsymbol{k}n}^{*} \tanh\!\left( \frac{E_{\boldsymbol{k}n}}{2T} \right) \cos k_{\mu}, \\[4pt]
\Delta_{\mu}^{t,\uparrow}
&= \frac{V}{2N} \sum_{\boldsymbol{k},n} i\, u_{1\boldsymbol{k}n} u_{3\boldsymbol{k}n}^{*} \tanh\!\left( \frac{E_{\boldsymbol{k}n}}{2T} \right) \sin k_{\mu}, \\[4pt]
\Delta_{\mu}^{t,\downarrow}
&= \frac{V}{2N} \sum_{\boldsymbol{k},n} i\, u_{2\boldsymbol{k}n} u_{4\boldsymbol{k}n}^{*} \tanh\!\left( \frac{E_{\boldsymbol{k}n}}{2T} \right) \sin k_{\mu}.
\end{aligned}
\label{eq:self_consistency}
\end{equation}

To characterize the boundary topology, we diagonalize the self-consistent BdG Hamiltonian in ribbon geometries. The spin-resolved local spectral function is evaluated from the retarded Green's function $G^{R}(k_{\mu},\omega) = [\omega + i\Gamma - \mathcal{H}_{\mathrm{BdG}}(k_{\mu})]^{-1}$ as
\begin{equation}
A_{\sigma}(m, k_{\mu}, \omega) = -\frac{1}{\pi} \operatorname{Im} G^{R}_{\alpha_{\sigma}(m), \alpha_{\sigma}(m)}(k_{\mu}, \omega),
\label{eq:spectral_function}
\end{equation}
where $k_{\mu}$ is the conserved momentum along the ribbon, $\Gamma$ is a small broadening factor, and $m$ labels the chain normal to the boundary. In the electron sector of the Nambu basis, $\alpha_{\uparrow}(m) = 2m-1$ and $\alpha_{\downarrow}(m) = 2m$. The boundary spectral functions are obtained by evaluating $A_{\sigma}(m, k_{\mu}, \omega)$ on the outermost chains.

In the following calculations, all energies are measured in units of $t_0$ (with $t_0 = 1$). Unless otherwise specified, we set $\mu=0$, $T = 10^{-6}$, a broadening factor $\Gamma = 0.006$ for the spectral functions, and the interaction strength $V = 1.5$.
	
	\section{Results and discussion}\label{sec:results}
 \begin{figure*}[t]
	\includegraphics[width=\textwidth]{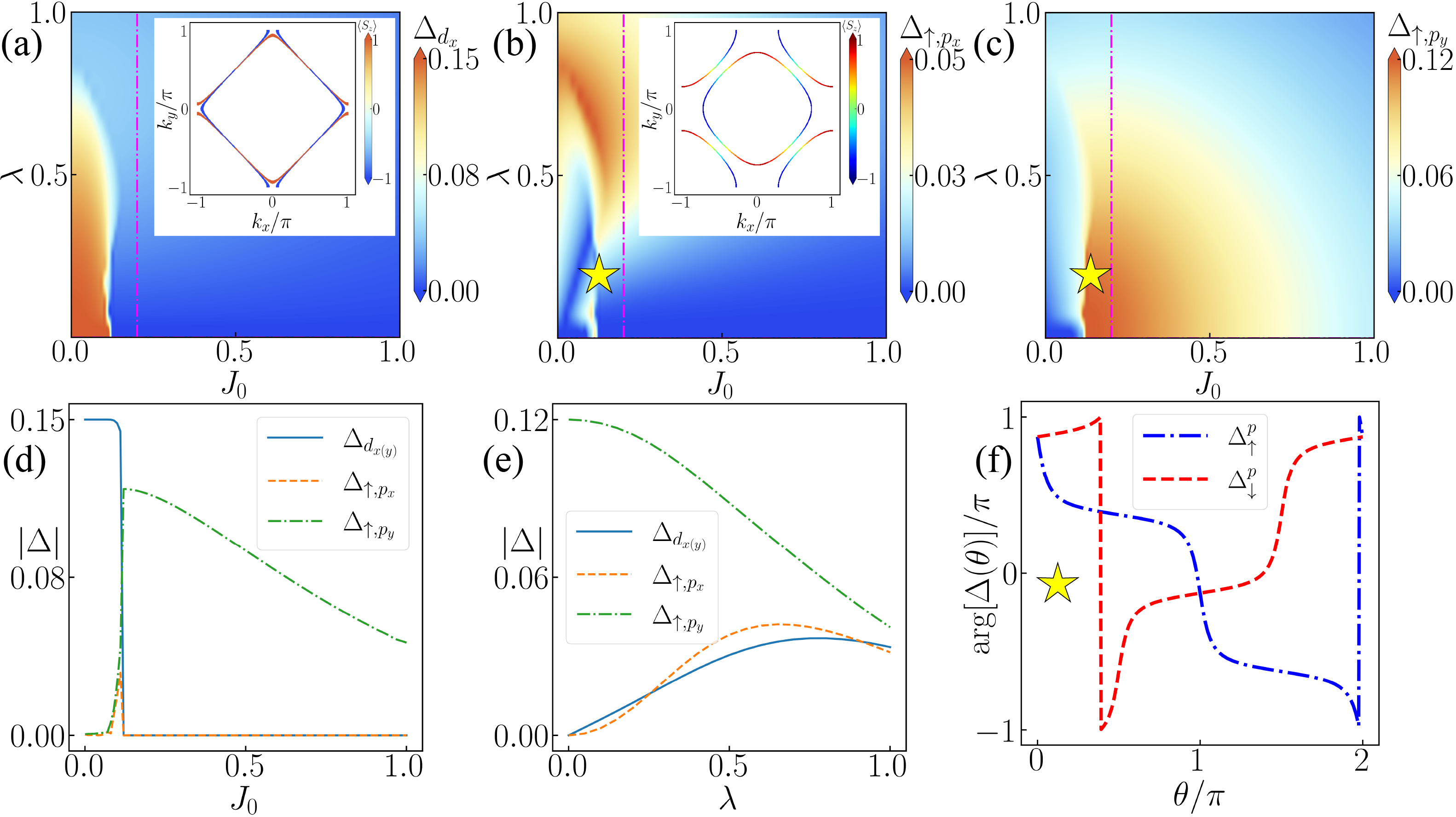}
	\caption{Self-consistent superconducting order parameters and their phase structure in the $d$-wave altermagnetic system with RSOC. (a)-(c) Magnitudes of the spin-singlet $d$-wave order parameter $\Delta_d$ and the equal-spin triplet components $\Delta_x^p$ and $\Delta_y^p$ as functions of the altermagnetic splitting $J_0$ and the RSOC strength $\lambda$. The insets show the Fermi surfaces for (a) $\lambda=0$, $J_0=0.01$ and (b) $\lambda=0.2$, $J_0=0.2$. (d) Order parameters as functions of $J_0$ at $\lambda=0$. (e) Order parameters as functions of $\lambda$ at fixed $J_0=0.2$. (f) Phase of the equal-spin triplet pairing along the Fermi surface for the spin-up and spin-down sectors at the star-marked point ($\lambda=0.2$, $J_0=0.12$).	
	}\label{fig:phase}
\end{figure*}

We begin by examining the superconducting order parameters obtained from the self-consistent calculation.
The self-consistent solutions reveal several robust features~\cite{supp}.
For the spin-singlet channel, we always obtain $\Delta_x^s = -\Delta_y^s$, indicating a dominant $d_{x^2-y^2}$-wave pairing (hereafter referred to as $d$-wave).
For the equal-spin triplet channel, the pairing is highly anisotropic.
Unlike previous studies that often impose predefined pairing symmetries (e.g., chiral $p_x+ip_y$) \cite{zhu_topological_2023}, our self-consistent calculation treats the $x$ and $y$ bond orders independently [Eq. \eqref{eq:self_consistency}], revealing a direction-selective triplet pairing.
In the absence of RSOC ($\lambda=0$), the triplet order becomes quasi-one-dimensional within each spin sector: for spin up, only the $y$-directed bond component is finite ($\Delta_y^{t,\uparrow}\neq0$, $\Delta_x^{t,\uparrow}\approx 0$); for spin down, the opposite holds ($\Delta_x^{t,\downarrow}\neq0$, $\Delta_y^{t,\downarrow} \approx 0$).
The preserved $\mathcal{C}_{4z}\mathcal{T}$ symmetry further relates the two spin sectors, enforcing $\Delta_y^{t,\uparrow} = \Delta_x^{t,\downarrow}$.
Thus, the triplet pairing (henceforth $p$-wave) in opposite spin sectors is linked by a $\pi/2$ rotation.
Consequently, when presenting the self-consistent pairing amplitudes, we may focus on the spin-up triplet components without loss of generality.

Figure~\ref{fig:phase} summarizes the self-consistent order parameters as functions of $J_0$ and $\lambda$.
Figures~\ref{fig:phase}(a)-\ref{fig:phase}(c) show the magnitudes of the $d$-wave singlet amplitude $\Delta_{d_x}$ and the spin-up triplet components $\Delta_{\uparrow,p_x}$ and $\Delta_{\uparrow,p_y}$ (where $\Delta_{\uparrow,p_\mu}\equiv\Delta_{\mu}^{t,\uparrow}$) over the $J_0$-$\lambda$ plane.
Figure~\ref{fig:phase}(d) displays the evolution with $J_0$ at $\lambda=0$, clearly demonstrating the emergence of $\Delta_{\uparrow,p_y}$ and $\Delta_{\downarrow,p_x}$ (the latter not shown) while $\Delta_{\uparrow,p_x}$ and $\Delta_{\downarrow,p_y}$ remain negligible.
Figure~\ref{fig:phase}(e) shows the dependence on $\lambda$ at fixed $J_0$, and Fig.~\ref{fig:phase}(f) illustrates the momentum-space phase winding of the triplet pairing at a representative point.

For small $J_0$, the altermagnetic spin splitting is modest, as seen in the Fermi surface inset of Fig.~\ref{fig:phase}(a); the Fermi-surface mismatch between opposite spins is limited, which favors singlet $d$-wave pairing.
Indeed, Fig.~\ref{fig:phase}(a) shows that the $d$-wave component dominates at weak $J_0$, while triplet components remain small.
With increasing $J_0$, the momentum-dependent spin splitting reconstructs the Fermi surface and enhances the spin-dependent mismatch, leading to a progressive suppression of singlet pairing and a relative enhancement of triplet pairing.

RSOC qualitatively modifies this pairing structure.
As indicated by the Fermi surface in the inset of Fig.~\ref{fig:phase}(b), the electronic states acquire mixed spin character.
This relaxes the spin-selective pairing pattern at $\lambda=0$ and activates otherwise suppressed pairing components, allowing singlet and triplet orders to coexist in a mixed-parity superconducting state.
Most notably, in each spin sector the originally quasi-one-dimensional triplet pairing acquires a finite $p_x$ component, evolving into a $p_x+ip_y$-like state, as evidenced by the finite $\Delta_{\uparrow,p_x}$ seen in Fig.~\ref{fig:phase}(b) at finite $\lambda$.
Nevertheless, the $p_y$ component remains dominant over the $p_x$ component, as can be seen by comparing $\Delta_{\uparrow,p_y}$ [Fig.~\ref{fig:phase}(c)] and $\Delta_{\uparrow,p_x}$ [Fig.~\ref{fig:phase}(b)], indicating that the induced $p+ip$ pairing retains a strong anisotropy inherited from the altermagnetic spin splitting.

To further examine the internal structure of the triplet order parameter, we focus on the $\lambda=0$ limit, where $s_z$ is conserved and the two spin sectors are decoupled.
In this limit, the altermagnetic exchange field lifts the equivalence between the two in-plane directions within each spin sector, allowing the triplet order to select a preferred bond direction.
As shown in Fig.~\ref{fig:phase}(d), increasing $J_0$ suppresses the singlet amplitude, while only two equal-spin triplet components become sizable: $\Delta_{\uparrow,p_y}$ (and, by $\mathcal{C}_{4z}\mathcal{T}$ symmetry, $\Delta_{\downarrow,p_x}$). 
In contrast, $\Delta_{\uparrow,p_x}$ and $\Delta_{\downarrow,p_y}$ remain strongly suppressed~\cite{supp}.
This demonstrates that altermagnetic spin splitting not only favors triplet pairing but also imposes a spin-dependent directional selection within the triplet sector.

The origin of this directional selectivity can be understood as follows.
From the perspective of the self-consistent pairing kernel [Eq. \eqref{OP:triplet}], the two triplet components in the spin-up sector can be written schematically as
\begin{equation}
	\Delta_{\mu\uparrow}^p \propto \sum_{\boldsymbol{k}} \mathcal{F}_{\uparrow\uparrow}(\boldsymbol{k}) \sin k_\mu , \qquad \mu=x,y.
	\label{eq:pwave_kernel}
\end{equation}
The relative magnitudes of $\Delta_{x\uparrow}^p$ and $\Delta_{y\uparrow}^p$ are governed by the projections of $\mathcal{F}_{\uparrow\uparrow}(\boldsymbol{k})$ onto the odd-parity form factors $\sin k_x$ and $\sin k_y$.
Because the altermagnetic exchange field has a $d_{x^2-y^2}$ character, it distinguishes the two in-plane directions within a fixed spin sector and yields an anisotropic pairing kernel, making the two projections unequal.
This selects one triplet bond component over the other.

The same phenomenon can be further understood within a phenomenological Ginzburg-Landau theory~\cite{supp}.
For a conventional $\mathcal{C}_{4z}$-symmetric two-component $p$-wave order parameter $\boldsymbol{\eta}=(\eta_x,\eta_y)$, the quadratic free energy is isotropic,
\[
F^{(2)} = \alpha \left(|\eta_x|^2 + |\eta_y|^2\right),
\]
so that $p_x$ and $p_y$ are degenerate at quadratic order.
In the altermagnetic state, the unitary $\mathcal{C}_{4z}$ symmetry is broken within each spin sector, allowing an anisotropic quadratic free energy.
For the spin-up sector, we may write
\[
F^{(2)}_\uparrow = \alpha_{x\uparrow} |\eta_{x\uparrow}|^2 + \alpha_{y\uparrow} |\eta_{y\uparrow}|^2,
\]
with $\alpha_{x\uparrow}\neq\alpha_{y\uparrow}$ in general.
If $\alpha_{y\uparrow}<\alpha_{x\uparrow}$, the leading instability occurs in the $p_y$ channel, giving $(\eta_{x\uparrow},\eta_{y\uparrow})\propto(0,1)$ (see Supplemental Material for details~\cite{supp}).
Thus, this symmetry-constrained quadratic anisotropy explains why $\Delta_{\uparrow,p_y}$ (and $\Delta_{\downarrow,p_x}$) dominate in the self-consistent solution, while $\Delta_{\uparrow,p_x}$ (and $\Delta_{\downarrow,p_y}$) remain strongly suppressed.

We next turn to the finite-Rashba-coupling regime.
Figure~\ref{fig:phase}(e) shows the evolution of the superconducting order parameters with $\lambda$ at fixed $J_0$.
As $\lambda$ increases, the triplet components suppressed in the $\lambda=0$ limit acquire finite amplitudes, accompanied by an enhancement of the $d$-wave component.
This behavior reflects Rashba-induced spin mixing, which relaxes the spin-selective pairing constraint and allows singlet and triplet components to admix.
Consequently, the superconducting state evolves from a direction-selective triplet state into a mixed-parity phase with coexisting pairing components.
The coexistence of the two triplet bond components also gives rise to a nontrivial phase texture in momentum space~\cite{supp}.
When both $p_x$ and $p_y$ components are finite, the triplet pairing function in Eq.~\eqref{eq:triplet_k} acquires a momentum-dependent complex phase.
We evaluate $\phi_\sigma(\boldsymbol{k})=\arg[\Delta^{p,\sigma}(\boldsymbol{k})]$ along a circular loop $\boldsymbol{k}(\theta)=R(\cos\theta,\sin\theta)$ centered at $\Gamma$ and define the corresponding winding as
\begin{equation}
	W_\sigma(R) = \frac{1}{2\pi} \int_0^{2\pi} d\theta \, \partial_\theta \phi_\sigma[\boldsymbol{k}(\theta)].
\end{equation}
As shown in Fig.~\ref{fig:phase}(f), the two spin sectors exhibit opposite windings, with $W_\uparrow=-1$ and $W_\downarrow=+1$.
This indicates that Rashba-induced spin mixing generates opposite momentum-space phase windings of the equal-spin triplet pairing in the two spin sectors.

	\begin{figure}[t]
		\includegraphics[scale=0.355]{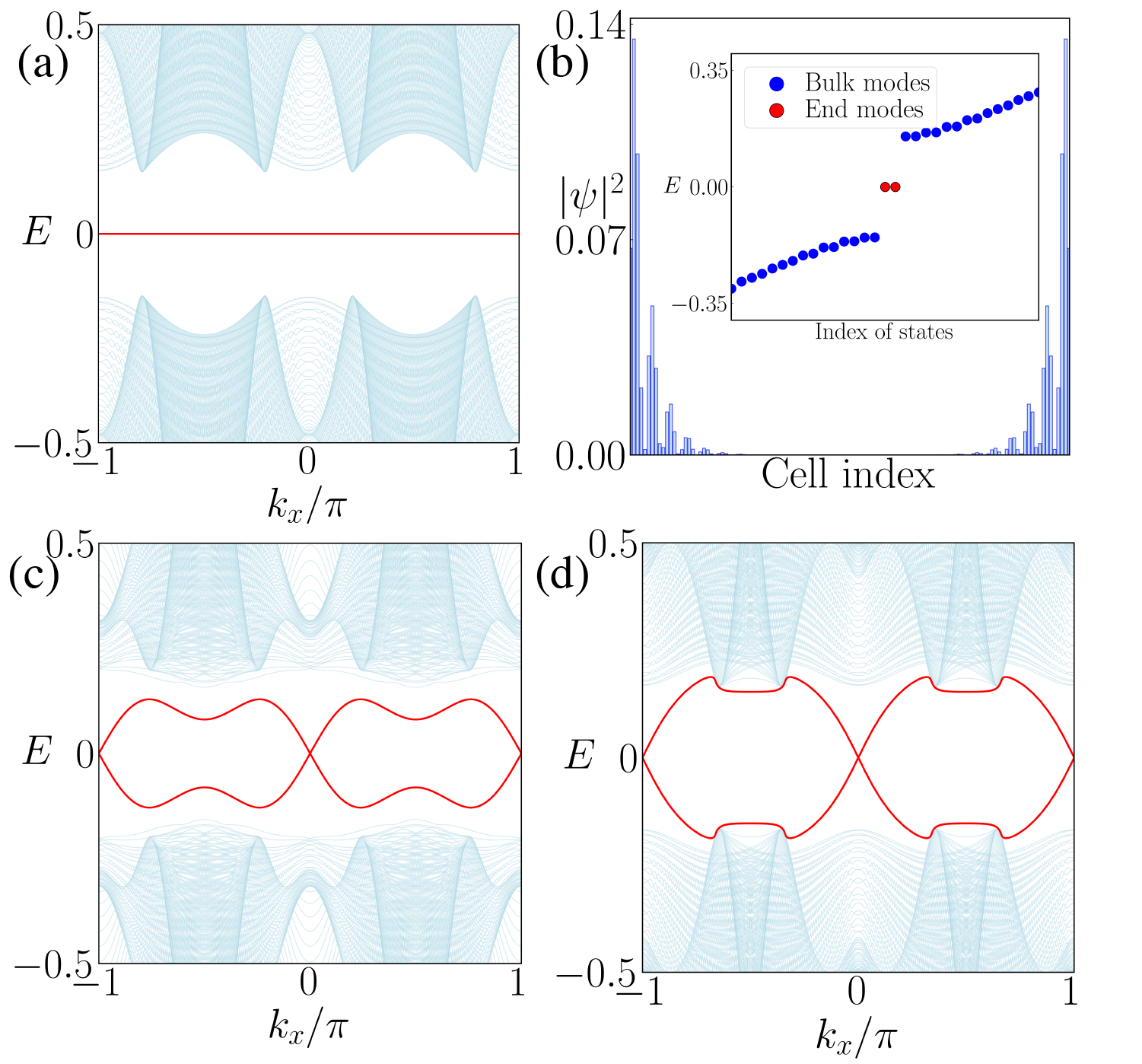}
		\caption{
Boundary spectra and zero-mode wave functions of the self-consistent superconducting phases with $J_0 = 0.12$.
(a) Cylinder spectrum at $\lambda = 0$.
(b) Energy levels and spatial distributions of the two zero-energy states at a representative momentum $k_x = 0$ selected from (a).
(c) Cylinder spectrum at $\lambda = 0.3$.
(d) Cylinder spectrum at $\lambda = 0.6$.
		}\label{fig:topo}
	\end{figure}
	
	\subsection{Boundary spectrum and topological properties}

The momentum-space phase winding of the triplet pairing motivates an analysis of the boundary spectrum. 
We therefore diagonalize the BdG Hamiltonian in ribbon geometries. 
We first consider the $\lambda=0$ limit. 
As discussed above, the self-consistent solution is dominated by two equal-spin triplet components, $\Delta^p_{y\uparrow}$ and $\Delta^p_{x\downarrow}$, whereas $\Delta^p_{x\uparrow}$ and $\Delta^p_{y\downarrow}$ remain negligible. 
Thus, each spin sector realizes $p$-wave pairing predominantly along a single crystalline direction, rather than a two-dimensional chiral $p_x\pm i p_y$ state. 
This anisotropic pairing structure effectively decomposes the system into momentum-resolved one-dimensional spinless $p$-wave superconducting channels~\cite{kitaev_unpaired_2001} and is directly reflected in the cylinder spectrum shown in Fig.~\ref{fig:topo}(a). 
For a fixed momentum along the periodic direction, the two-dimensional BdG Hamiltonian reduces to an effective one-dimensional problem along the open direction. 
Over a finite momentum interval, this effective one-dimensional subsystem supports Majorana end modes at the two open boundaries. 
These momentum-resolved end states appear as nearly dispersionless zero-energy boundary states in the cylinder spectrum.

\begin{figure}[t]
		\includegraphics[scale=0.36]{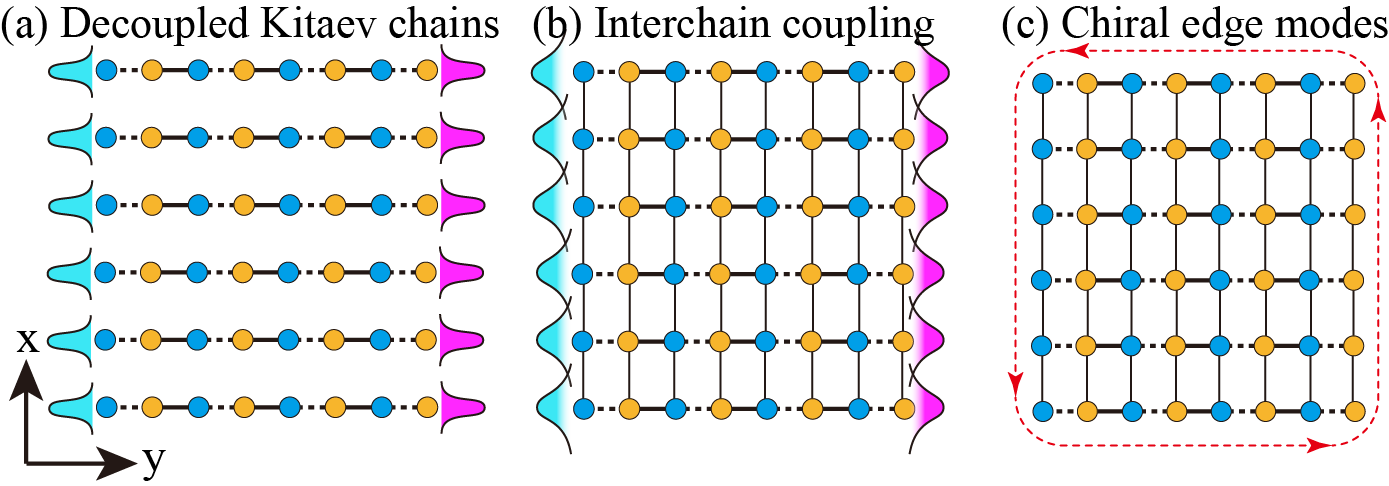}
		\caption{
			Schematic illustration of the evolution from decoupled Kitaev chains to a two-dimensional Majorana edge mode.
(a) Decoupled Kitaev chains stacked along the $x$ direction, each hosting Majorana zero modes at its two ends.
(b) Interchain coupling hybridizes the boundary Majorana modes.
(c) In the coupled two-dimensional limit, the hybridized Majorana modes form a dispersive edge channel along the boundary.
		}
		\label{fig:kitaev}
	\end{figure}

	To verify the boundary localization, we select a representative momentum $k_x$ from the zero-energy manifold and diagonalize the corresponding finite one-dimensional BdG Hamiltonian. 
	As shown in Fig.~\ref{fig:topo}(b), the two zero-energy eigenstates are localized at opposite open boundaries, consistent with Majorana end modes.	This behavior has a simple real-space interpretation, as illustrated in Fig.~\ref{fig:kitaev}(a). In the $\lambda=0$ limit, the anisotropic pairing effectively decomposes the system into decoupled Kitaev-chain-like channels stacked along the x direction. 
	Each channel supports Majorana zero modes at its ends, and their momentum-resolved collection gives rise to the nearly flat zero-energy boundary states in Fig.~\ref{fig:topo}(a).
	
	With finite RSOC, the boundary spectrum changes qualitatively, as shown in Fig.~\ref{fig:topo}(c). 
	The nearly flat zero-energy modes become dispersive away from the high-symmetry momenta, while gapless crossings remain at $k_x=0$ and $k_x=\pi$. 
	This evolution reflects Rashba-induced spin mixing, which couples the previously decoupled spin sectors and allows both $p_x$ and $p_y$ pairing components to coexist.	In the real-space picture, finite RSOC introduces coupling between the Kitaev-chain-like channels, as illustrated in Fig.~\ref{fig:kitaev}(b). 
	The resulting hybridization of boundary Majorana modes converts the flat zero-energy states into dispersive edge modes of the two-dimensional superconducting state, as shown in Fig.~\ref{fig:kitaev}(c). The residual gapless crossings at $k_x=0$ and $k_x=\pi$ can be traced to the vanishing of the Rashba term proportional to $\sin k_x$ at these momenta. 
	The suppressed spin mixing limits the hybridization of boundary states and allows the crossings to persist. For stronger RSOC, the hybridization between boundary and bulk states is further enhanced, as shown in Fig.~\ref{fig:topo}(d). 
	The edge modes gradually lose their spectral separation from the bulk continuum. 
	To clarify the topological character of these boundary features, we evaluate the bulk invariant of the two-dimensional BdG Hamiltonian.
	
\begin{figure}[t]
		\includegraphics[width=\columnwidth]{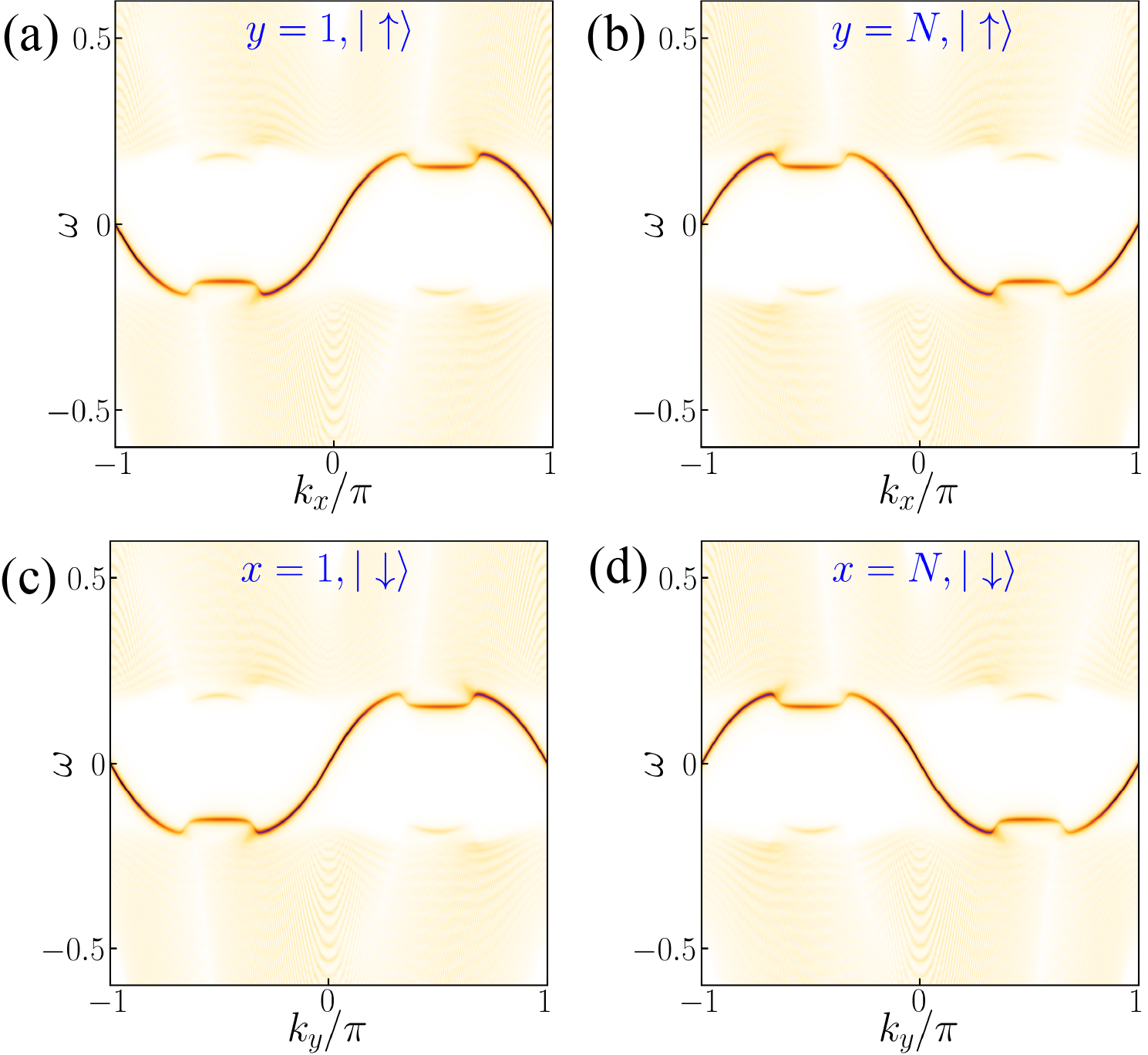}
		\caption{
			Spin-resolved boundary spectral functions calculated from the Green's function with $J_0=0.12$ and $\lambda=0.6$. 
			(a) Spin-up spectral weight at the lower edge ($y=1$). 
			(b) Spin-up spectral weight at the upper edge ($y=N$). 
			(c) Spin-down spectral weight at the left edge ($x=1$). 
			(d) Spin-down spectral weight at the right edge ($x=N$). 
			The boundary states exhibit clear dispersion, with opposite slopes near $k=0$ and $k=\pi$. 
		}\label{fig:edge}
	\end{figure}

	The superconducting state belongs to class D because time-reversal symmetry is broken by the altermagnetic order~\cite{schnyder_classification_2008}. 
	In two dimensions, a nonzero Chern number in this class would correspond to a chiral topological superconducting phase with a net chiral Majorana edge channel. 
	Our calculation, however, yields a vanishing total Chern number. 
	This is consistent with the boundary spectrum, where the edge branches near $k_x=0$ and $k_x=\pi$ disperse with opposite chiralities and therefore do not produce a net chiral edge response [Fig.~\ref{fig:edge}].
	
	The residual gapless crossings nevertheless admit a natural momentum-resolved interpretation. 
	At $k_x=0$ and $k_x=\pi$, the Rashba term proportional to $\sin k_x$ vanishes, thereby suppressing the hybridization between the relevant boundary channels. 
	At these fixed momenta, the BdG Hamiltonian can be viewed as an effective one-dimensional $p$-wave superconductor along the open direction, which supports Majorana end states. The gapless crossings in the ribbon spectrum can therefore be interpreted as the momentum-resolved spectral signatures of these one-dimensional Majorana end states~\cite{zhu_topological_2023,Alam_prox_2026}.
	
	\subsection{Boundary orientation and spin polarization}
	Motivated by the symmetry properties of altermagnets, where spin degeneracy is lifted in a momentum-dependent manner while the net magnetization remains zero, we characterize the spin structure of the boundary states through spin-resolved spectral functions. 
	The spin polarization depends sensitively on both momentum and boundary orientation, making a spin-resolved analysis essential. 
	We therefore evaluate the boundary spectral functions using the retarded Green's function defined in Eq.~\eqref{eq:spectral_function}. Figures~\ref{fig:edge}(a)-\ref{fig:edge}(d) show representative spectral weights at the four boundaries of the ribbon geometry. 
	For clarity, only the dominant spin component at each boundary is displayed; the complete set of spin-resolved spectra for all boundary-spin combinations is provided in the Supplemental Material~\cite{supp}. 
	
	For the boundaries at $y=1$ and $y=N$, the low-energy spectral weight is dominated by the spin-up component, whereas for the boundaries at $x=1$ and $x=N$, the dominant contribution arises from the spin-down component. 
	Thus, changing the boundary orientation from the $y$ edge to the $x$ edge switches the dominant spin sector of the boundary states. 
	This establishes a clear correlation between boundary orientation and spin polarization, reflecting the momentum-dependent spin splitting of the altermagnetic normal state rather than conventional Rashba spin-momentum locking.
	
	This behavior follows directly from the $\mathcal{C}_{4z}\mathcal{T}$ symmetry of the altermagnetic state. 
	Under this symmetry, a $\pi/2$ rotation in real space is accompanied by spin reversal, mapping the boundary states on the $y$ edges to those on the $x$ edges with opposite dominant spin polarization. 
	Consequently, the boundary orientation determines the dominant spin character of the boundary states.
	Such a spin-edge locking provides a characteristic signature of altermagnetic superconductivity and may be probed by spin-resolved tunneling spectroscopy~\cite{wiesendangerSpin2009} or spin-sensitive quasiparticle-interference measurements~\cite{hoffmanSpectroscopic2011}.
	
	\section{Summary}\label{sec:summary}
	In summary, we have investigated the influence of $d$-wave altermagnetic spin splitting on superconducting pairing and boundary properties in a two-dimensional metal. 
	Self-consistent calculations demonstrate that altermagnetism suppresses singlet pairing and stabilizes anisotropic equal-spin triplet order. 
	Within a Ginzburg-Landau framework, this behavior is attributed to a symmetry-allowed quadratic anisotropy between the $p_x$ and $p_y$ channels, which selects direction-dependent triplet components, while the $\mathcal{C}_{4z}\mathcal{T}$ symmetry constrains their relation between opposite spin sectors. 	RSOC further mixes the spin sectors and activates additional pairing channels, resulting in a mixed-parity superconducting state with nontrivial triplet phase winding. 
	This modification is reflected in the boundary spectrum, which evolves from nearly flat Majorana modes associated with effectively one-dimensional channels to dispersive boundary excitations. 
	
	The spin-resolved boundary spectra reveal a characteristic spin-edge locking, in which the dominant spin polarization of the boundary states is determined by the crystallographic orientation of the edge. 
	This feature follows from the underlying $\mathcal{C}_{4z}\mathcal{T}$ symmetry and provides a distinct spectroscopic signature. 
	These results establish altermagnetic metals as a platform for symmetry-controlled pairing and spin-selective Majorana boundary phenomena without external magnetic fields.


%

\renewcommand{\thesection}{S-\arabic{section}}
\setcounter{section}{0}  
\renewcommand{\theequation}{S\arabic{equation}}
\setcounter{equation}{0}  
\renewcommand{\thefigure}{S\arabic{figure}}
\setcounter{figure}{0}  
\renewcommand{\thetable}{S\Roman{table}}
\setcounter{table}{0}  
\onecolumngrid \flushbottom 
\newpage

\begin{center}\large \textbf{Supplementary material for “Direction-selective triplet pairing and spin-edge locking in altermagnetic metals"}\end{center}
	
In this Supplementary Material, we provide additional theoretical analysis and supporting results for the main text. 
Section~\ref{sec:GL} presents a Ginzburg--Landau analysis of the two-component $p$-wave order parameter and shows how altermagnetic anisotropy leads to direction-selective triplet pairing. 
Section~\ref{sec:edge} gives a complete account of the self-consistent superconducting state, including the pairing amplitudes in both spin sectors, their symmetry relations, the momentum-space phase winding of the triplet pairing, and the spin-resolved boundary spectral functions underlying the spin--edge correspondence.

\section{Ginzburg--Landau analysis of altermagnetic anisotropy}\label{sec:GL}

We analyze the effect of altermagnetic anisotropy on the equal-spin triplet order parameter within a Ginzburg--Landau (GL) theory~\cite{sigristPhenomenological1991}. 
For a two-component $p$-wave order parameter, the superconducting gap can be written as
\begin{equation}
	\Delta(\boldsymbol{k})
	=
	\eta_x \varphi_x(\boldsymbol{k})
	+
	\eta_y \varphi_y(\boldsymbol{k}),
\end{equation}
where $\eta_x$ and $\eta_y$ are complex order-parameter components and the basis functions may be chosen as $\varphi_x(\boldsymbol{k})\sim \sin k_x$ and $\varphi_y(\boldsymbol{k})\sim \sin k_y$. 
The GL free energy is expanded as
\begin{equation}
	F
	=
	F^{(2)}
	+
	F^{(4)}
	+
	\cdots ,
\end{equation}
with
\begin{equation}
	F^{(2)}
	=
	\sum_{ij}
	\alpha_{ij}
	\eta_i^*
	\eta_j ,
	\qquad
	F^{(4)}
	=
	\sum_{ijkl}
	\beta_{ijkl}
	\eta_i^*
	\eta_j^*
	\eta_k
	\eta_l .
\end{equation}
Here, the quadratic coefficients determine the leading superconducting instability, whereas the quartic terms determine how different components compete or coexist once the order parameter becomes finite.

We first recall the conventional $C_4$-symmetric case for comparison. 
In this case, the two components $p_x$ and $p_y$ form a two-dimensional representation. 
The quadratic term is constrained to be isotropic,
\begin{equation}
	F^{(2)}_{\rm iso}
	=
	\alpha
	\left(
	|\eta_x|^2
	+
	|\eta_y|^2
	\right),
\end{equation}
so that $p_x$ and $p_y$ remain degenerate at quadratic order. 
The relative amplitude and phase are therefore selected by quartic invariants. 
A convenient form of the quartic free energy is
\begin{equation}
	F^{(4)}_{\rm iso}
	=
	\beta_1
	\left(
	|\eta_x|^2
	+
	|\eta_y|^2
	\right)^2
	+
	\beta_2
	\left|
	\eta_x^2
	+
	\eta_y^2
	\right|^2
	+
	\beta_3
	|\eta_x|^2
	|\eta_y|^2 .
\end{equation}
The $\beta_2$ term controls the relative phase between the two components. 
For $\beta_2>0$, the free energy favors minimizing $|\eta_x^2+\eta_y^2|^2$, which is achieved by the chiral combinations
\begin{equation}
	(\eta_x,\eta_y)
	\propto
	(1,\pm i).
\end{equation}
For $\beta_2<0$, the free energy instead favors real combinations. 
Within the real sector, the $\beta_3$ term further distinguishes whether a single-component state, such as $(1,0)$ or $(0,1)$, or a two-component real state, such as $(1,\pm 1)$, is preferred. 
Thus, in a $C_4$-symmetric system, $p_x$ and $p_y$ are degenerate at quadratic order, and the detailed pairing structure is selected by quartic terms.

In the present model, the unitary $C_{4z}$ symmetry and time-reversal symmetry $\mathcal{T}$ are separately broken, while the combined symmetry $C_{4z}\mathcal{T}$ is preserved. 
Consequently, within a fixed spin sector, $p_x$ and $p_y$ are no longer related by a unitary fourfold rotation and need not have the same quadratic coefficient. 
The quadratic GL free energy for one spin sector can therefore contain an anisotropic term,
\begin{equation}
	F^{(2)}_{\rm AM}
	=
	\alpha_x
	|\eta_x|^2
	+
	\alpha_y
	|\eta_y|^2 ,
\end{equation}
with $\alpha_x\neq\alpha_y$ in general. 
Equivalently, one may write
\begin{equation}
	\alpha_x
	=
	\alpha
	+
	\delta\alpha ,
	\qquad
	\alpha_y
	=
	\alpha
	-
	\delta\alpha ,
\end{equation}
so that $\delta\alpha$ measures the altermagnetic anisotropy between the two $p$-wave components. This anisotropy changes the mechanism of order-parameter selection. 
If $\delta\alpha<0$, then $\alpha_x<\alpha_y$, and the $p_x$ component has the stronger superconducting instability. 
The leading state is then
\begin{equation}
	(\eta_x,\eta_y)
	\propto
	(1,0).
\end{equation}
Conversely, if $\delta\alpha>0$, the $p_y$ component is favored and
\begin{equation}
	(\eta_x,\eta_y)
	\propto
	(0,1).
\end{equation}
Therefore, unlike the isotropic $C_4$-symmetric case, the altermagnetic state can select a single $p$-wave component already at the quadratic level. 
The quartic terms still affect possible subleading admixtures and the stability of two-component states, but the leading directional selection is controlled by the anisotropic quadratic coefficients.

For the full spinful altermagnetic system, the remaining $C_{4z}\mathcal{T}$ symmetry relates the two spin sectors. 
This symmetry maps a spin-up component with $p_y$ character to a spin-down component with $p_x$ character, and similarly maps a spin-up component with $p_x$ character to a spin-down component with $p_y$ character. 
Accordingly, the quadratic coefficients obey
\begin{equation}
	\alpha_{y\uparrow}
	=
	\alpha_{x\downarrow},
	\qquad
	\alpha_{x\uparrow}
	=
	\alpha_{y\downarrow}.
\end{equation}
Thus, if the spin-up sector favors $\eta_{y\uparrow}$, the spin-down sector favors $\eta_{x\downarrow}$. 
The resulting leading pairing structure is
\begin{equation}
	\eta_{y\uparrow}\neq 0,
	\qquad
	\eta_{x\downarrow}\neq 0,
	\qquad
	\eta_{x\uparrow}\simeq 0,
	\qquad
	\eta_{y\downarrow}\simeq 0 .
\end{equation}
This is in direct agreement with the self-consistent solution, where $\Delta^p_{y\uparrow}$ and $\Delta^p_{x\downarrow}$ are the dominant equal-spin triplet components. 
The essential point is that the directional selection in the altermagnetic state is not primarily determined by the sign of a quartic coefficient such as $\beta_2$, but by the quadratic anisotropy induced by the $d_{x^2-y^2}$-type altermagnetic spin splitting.

\section{Complete self-consistent pairing structure}\label{sec:edge}

\subsection{Pairing amplitudes and $C_{4z}\mathcal{T}$ symmetry}

In the main text, we show representative self-consistent order parameters to illustrate the evolution of the superconducting state. 
For completeness, here we present the full set of pairing components in both spin sectors. 
This analysis provides a direct check of how the self-consistent pairing structure is constrained by the combined $C_{4z}\mathcal{T}$ symmetry of the altermagnetic state.

We first consider the case without Rashba spin-orbit coupling, $\lambda=0$. 
In this limit, $s_z$ is conserved and the two spin sectors are decoupled. 
We extract the equal-spin triplet components $\Delta^p_{x\sigma}$ and $\Delta^p_{y\sigma}$ for $\sigma=\uparrow,\downarrow$, together with the opposite-spin $d$-wave singlet component $\Delta_d$. 
The full results are shown in Figs.~\ref{fig:pair_val}(a) and \ref{fig:pair_val}(b). 
For the spin-up sector, the dominant triplet component is $\Delta_{\uparrow,p_y}$, whereas $\Delta_{\uparrow,p_x}$ remains strongly suppressed. 
For the spin-down sector, the dominant component is $\Delta_{\downarrow,p_x}$, while $	\Delta_{\downarrow,p_y}$ is negligible. 
Thus, the anisotropic triplet pairing is selected along opposite crystalline directions in the two spin sectors, as dictated by the combined $C_{4z}\mathcal{T}$ symmetry. The operation rotates the momentum structure by $\pi/2$ and flips the spin, thereby relating
\begin{equation}
	\Delta_{\uparrow,p_x}
	\longleftrightarrow
	\Delta_{\downarrow,p_y},
	\qquad
	\Delta_{\uparrow,p_y}
	\longleftrightarrow
		\Delta_{\downarrow,p_x}.
\end{equation}

We next examine the finite-Rashba-coupling case. 
For $\lambda\neq0$, spin is no longer a good quantum number because the Rashba term mixes the two spin sectors. 
Consequently, pairing components that are strongly suppressed at $\lambda=0$, such as $\Delta_{\uparrow,p_x}$ and $\Delta_{\downarrow,p_y}$, acquire finite amplitudes. 
The corresponding complete results are shown in Figs.~\ref{fig:pair_val}(c) and \ref{fig:pair_val}(d). 
Although Rashba spin--orbit coupling activates additional components, the pairing structure remains anisotropic: the components connected to the dominant $\lambda=0$ channels remain larger than the previously suppressed ones. 
This demonstrates that the altermagnetic anisotropy persists even after spin mixing is introduced.

The complete self-consistent order parameters therefore support the symmetry-based interpretation used in the main text. 
In the spin-decoupled limit, $C_{4z}\mathcal{T}$ relates the dominant components $\Delta_{\uparrow,p_y}$ and $\Delta_{\downarrow,p_x}$. 
With finite Rashba spin--orbit coupling, additional components appear, but the hierarchy of the triplet order parameters continues to reflect the underlying altermagnetic anisotropy.
\renewcommand \thefigure {S\arabic{figure}}
\begin{figure}[t]
	\centering
	\includegraphics[width=\textwidth]{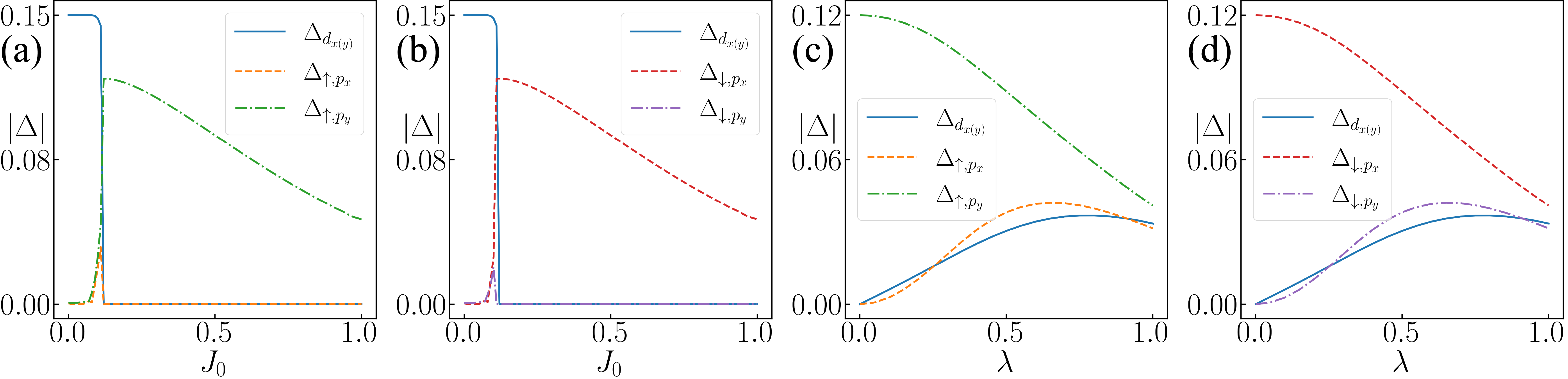}
	\caption{
		Complete self-consistent superconducting order parameters in the two spin sectors. 
		(a),(b) Order parameters as functions of $J_0$ at $\lambda=0$ for the spin-up and spin-down sectors, respectively. 
		(c),(d) Order parameters as functions of $\lambda$ at fixed $J_0=0.2$ for the spin-up and spin-down sectors, respectively. 
		The equal-spin triplet components $\Delta^p_{x\sigma}$ and $\Delta^p_{y\sigma}$ and the opposite-spin $d$-wave singlet component $\Delta_d$ are shown.
	}\label{fig:pair_val}		
\end{figure}

\subsection{Momentum-space phase structure of the triplet pairing}

We further analyze the momentum-space phase structure of the equal-spin triplet pairing. 
For a given spin sector, the triplet pairing function is written as
\begin{equation}
	\Delta^{p}_{\sigma}(\boldsymbol{k})
	=
	\Delta^{p,\sigma}_x \sin k_x
	+
	\Delta^{p,\sigma}_y \sin k_y ,
\end{equation}
and its phase is defined by
\begin{equation}
	\phi_\sigma(\boldsymbol{k})
	=
	\arg\!\left[
	\Delta^{p}_{\sigma}(\boldsymbol{k})
	\right].
\end{equation}

\begin{figure}[b]
	\centering
	\includegraphics[scale=0.62]{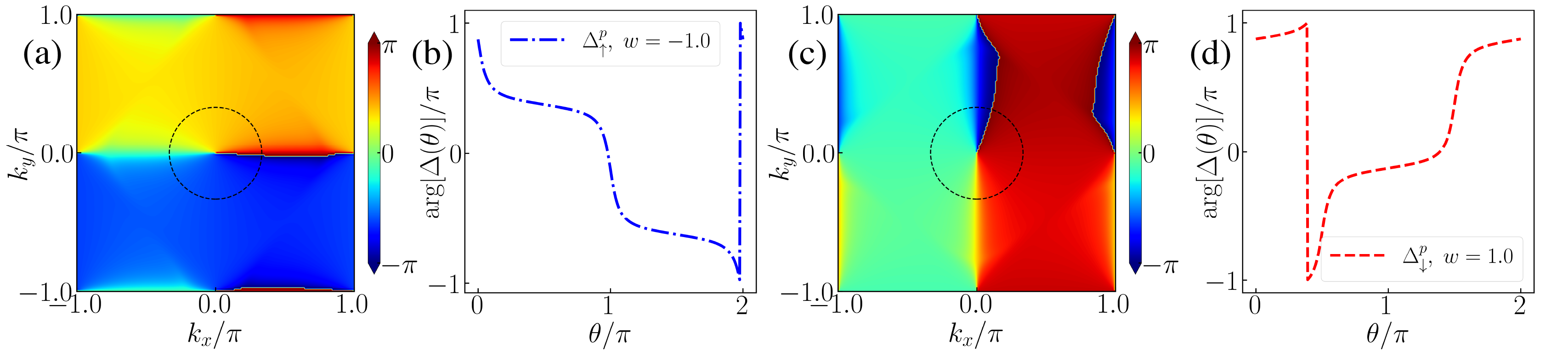}
	\caption{
		Momentum-space phase structure of the equal-spin triplet pairing. 
		(a),(c) Phase distribution $\phi_\sigma(\boldsymbol{k})$ in the Brillouin zone for the spin-up and spin-down triplet pairing functions, respectively. 
		(b),(d) Phase $\phi_\sigma(\theta)$ evaluated along a circular loop $\boldsymbol{k}(\theta)=R(\cos\theta,\sin\theta)$ centered at $\Gamma$ for the spin-up and spin-down sectors, respectively. 
		Common parameters are $J_0=0.12$ and $\lambda=0.6$.
	}\label{fig:pair_phase}
\end{figure}

Figures~\ref{fig:pair_phase}(a) and \ref{fig:pair_phase}(c) show the phase distribution $\phi_\sigma(\boldsymbol{k})$ in the Brillouin zone for the spin-up and spin-down triplet pairing functions, respectively. 
The color scale denotes the phase angle in the range $[-\pi,\pi]$. 
The phase exhibits strong momentum dependence, indicating a nontrivial internal phase structure of the equal-spin triplet pairing. 

To quantify this phase texture, we evaluate the phase winding along a closed loop in momentum space. 
The loop is chosen as a circle centered at $\Gamma$,
\begin{equation}
	\boldsymbol{k}(\theta)
	=
	k_0
	(\cos\theta,\sin\theta),
	\qquad
	0\leq\theta<2\pi ,
\end{equation}
where $k_0$ is a fixed radius chosen to avoid nodes of $\Delta^p_\sigma(\boldsymbol{k})$. 
Along this path, the phase becomes
\begin{equation}
	\phi_\sigma(\theta)
	=
	\arg\!\left[
	\Delta^{p}_{\sigma}(\boldsymbol{k}(\theta))
	\right].
\end{equation}
The corresponding winding number is
\begin{equation}
	W_\sigma
	=
	\frac{1}{2\pi}
	\int_0^{2\pi}
	d\theta\,
	\partial_\theta
	\phi_\sigma(\theta).
\end{equation}
The resulting $\phi_\sigma(\theta)$ is shown in Figs.~\ref{fig:pair_phase}(b) and \ref{fig:pair_phase}(d). 
As $\theta$ evolves from $0$ to $2\pi$, the phase winds by approximately $\pm 2\pi$, giving a nonzero winding number. 

\subsection{Spin-resolved boundary spectral functions}

\begin{figure}[h]
	\centering
	\includegraphics[width=\textwidth]{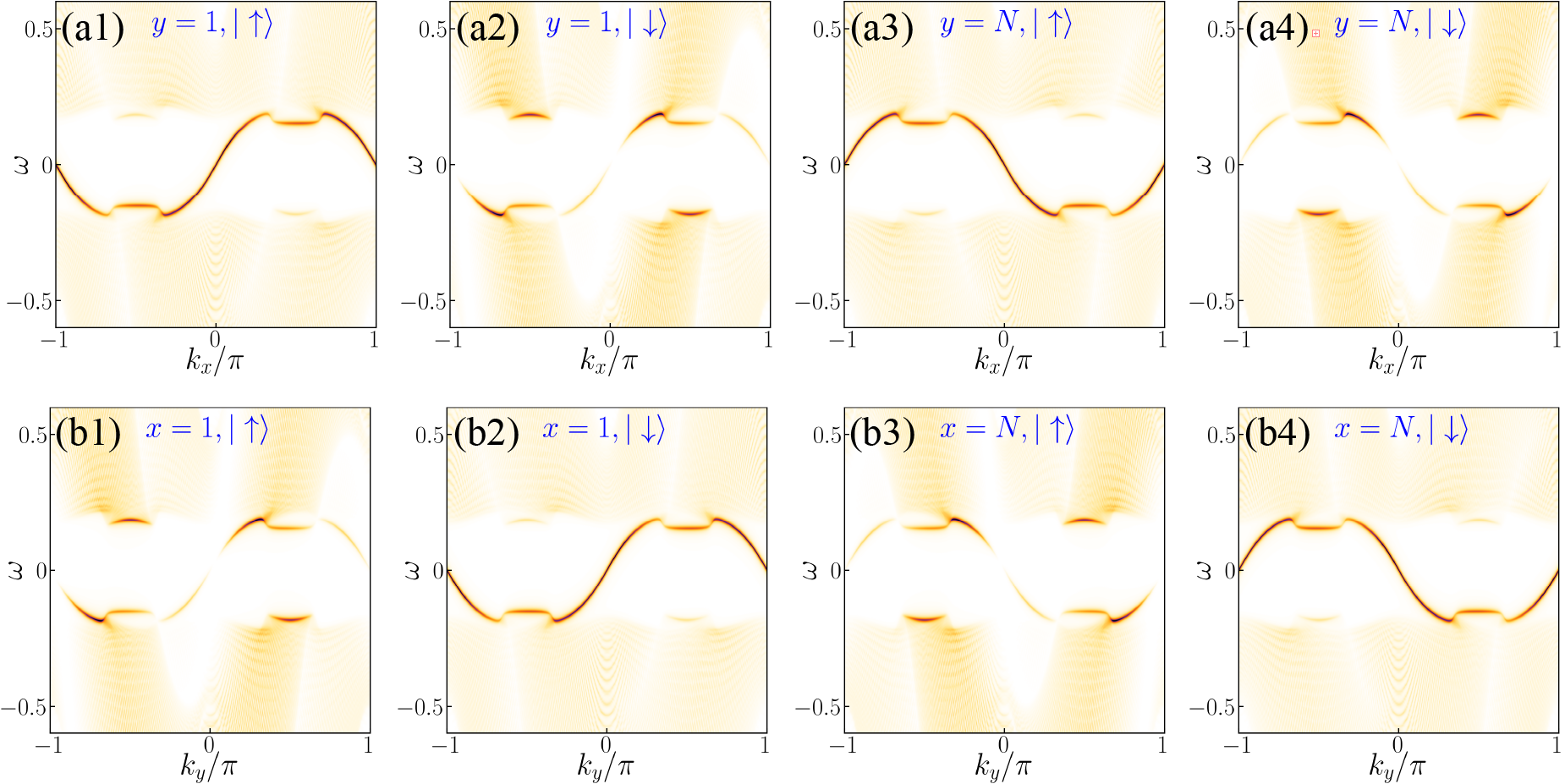}
	\caption{
		Complete spin-resolved boundary spectral functions. 
		(a1),(a2) Spin-up and spin-down spectral weights at the lower edge ($y=1$), respectively. 
		(a3),(a4) Spin-up and spin-down spectral weights at the upper edge ($y=N$), respectively. 
		(b1),(b2) Spin-up and spin-down spectral weights at the left edge ($x=1$), respectively. 
		(b3),(b4) Spin-up and spin-down spectral weights at the right edge ($x=N$), respectively. 
		Common parameters are $J_0=0.12$ and $\lambda=0.6$.
	}\label{fig:edge_spec}   
\end{figure}

In the main text, we present representative spin-resolved boundary spectra to illustrate the spin-edge correspondence of the superconducting state. 
For completeness, here we show the full set of boundary spectral functions, including both spin components on all four edges of the ribbon geometry. 
The spin-resolved boundary spectral function is evaluated from the retarded Green's function of the Bogoliubov-de Gennes Hamiltonian,
\begin{equation}
	G^R(k_\parallel,\omega)
	=
	\left[
	\omega + i\Gamma - \mathcal{H}_{\rm BdG}(k_\parallel)
	\right]^{-1},
\end{equation}
where $k_\parallel$ denotes the conserved momentum along the periodic direction and $\Gamma$ is a small broadening parameter. 
The local spectral function for spin $\sigma$ on the $m$-th chain is given by
\begin{equation}
	A_{\sigma}(m,k_\parallel,\omega)
	=
	-\frac{1}{\pi}
	\mathrm{Im}\,
	G^R_{\alpha_\sigma(m),\alpha_\sigma(m)}(k_\parallel,\omega),
\end{equation}
where $\alpha_\uparrow(m)=2m-1$ and $\alpha_\downarrow(m)=2m$ label the electron sector of the Nambu basis.

For each boundary, we evaluate the spectral function at the outermost chains, i.e., $m=1$ or $m=N$, and resolve the two spin components separately. 
The complete results are shown in Fig.~\ref{fig:edge_spec}, where the spectral weights are plotted for the four boundaries at $y=1$, $y=N$, $x=1$, and $x=N$ for both spin-up and spin-down sectors. 
The full data confirm a pronounced correlation between boundary orientation and spin polarization. 
For the boundaries normal to the $y$ direction ($y=1$ and $y=N$), the low-energy spectral weight is dominated by the spin-up component, while the spin-down contribution remains strongly suppressed. 
In contrast, for the boundaries normal to the $x$ direction ($x=1$ and $x=N$), the dominant spectral weight arises from the spin-down component. 
Thus, changing the boundary orientation from the $y$ edge to the $x$ edge switches the dominant spin sector of the boundary states.

\begin{figure}
\centering
\includegraphics[width=\textwidth]{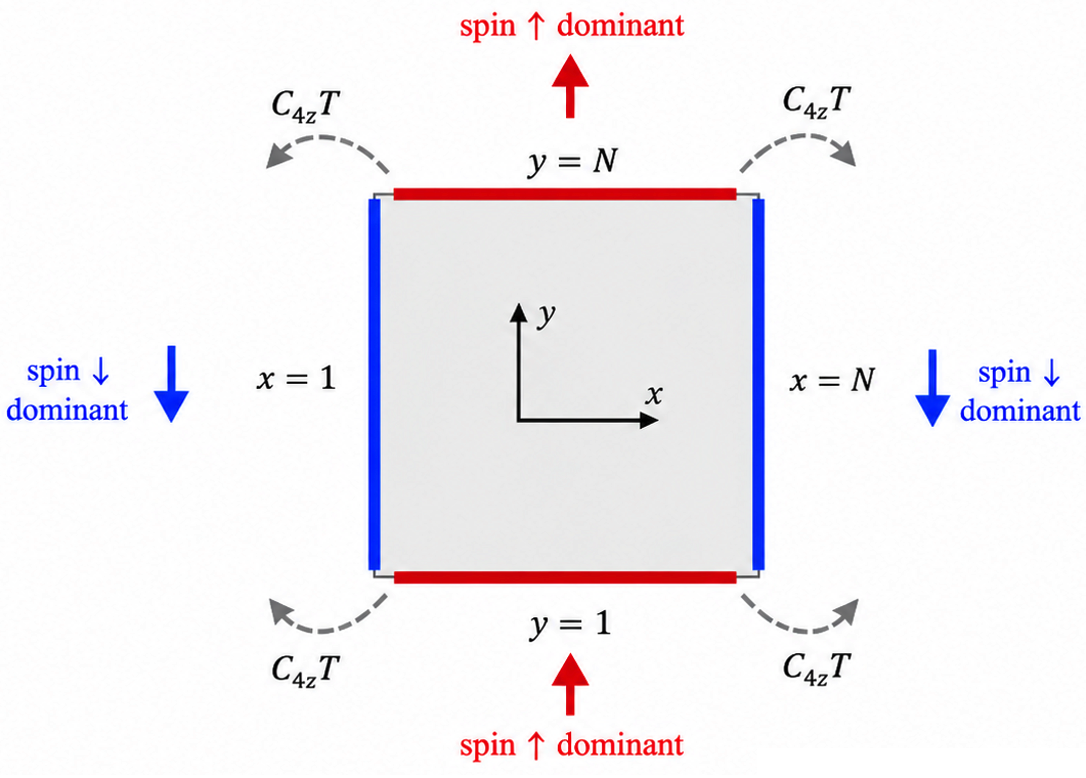}
\caption{
	Schematic illustration of the spin--edge correspondence in the altermagnetic superconducting state. 
	The boundaries normal to the $y$ direction ($y=1,N$) host spin-up dominated modes, while those normal to the $x$ direction ($x=1,N$) host spin-down dominated modes. 
	This edge-dependent spin polarization reflects the underlying $C_{4z}\mathcal{T}$ symmetry of the altermagnetic state.
}\label{fig:spin_edge_schematic}
\end{figure}

The subdominant components remain finite due to Rashba spin--orbit coupling, but their spectral weights are much smaller than those of the dominant components. 
The boundary states therefore retain a well-defined spin character on each edge. 
This spin--edge correspondence is summarized schematically in Fig.~\ref{fig:spin_edge_schematic}. 
The boundaries normal to the $y$ direction are dominated by spin-up spectral weight, whereas those normal to the $x$ direction are dominated by spin-down spectral weight.

This edge-dependent spin polarization follows from the $C_{4z}\mathcal{T}$ symmetry of the altermagnetic state. 
The $C_{4z}$ operation exchanges the $x$ and $y$ directions, while $\mathcal{T}$ reverses the spin, leading to the correspondence
\begin{equation}
	(y\text{-edge},\uparrow)
	\longleftrightarrow
	(x\text{-edge},\downarrow).
\end{equation}
Thus, boundary states on orthogonal edges are symmetry-related and carry opposite dominant spin components. 
The complete spin-resolved boundary spectra, together with the schematic in Fig.~\ref{fig:spin_edge_schematic}, provide a comprehensive confirmation of the spin--edge correspondence discussed in the main text.

\end{document}